\documentclass[amsfonts,amsmath,prd,preprint,nofootinbib]{revtex4}
\newcommand{\beq}{\begin{equation}}
\newcommand{\eeq}{\end{equation}}

\newcommand{\bsp}{\begin{split}}

\usepackage{epsfig,bbm,cancel,ulem}
\usepackage[breaklinks=true]{hyperref}
\usepackage{xcolor}
\usepackage{wasysym}
\usepackage[mathscr]{eucal}
\usepackage{multirow}
\usepackage{amsmath}
\usepackage{enumitem}

\begin{document}

\title{Strong lensing and shadow of Ayon-Beato-Garcia (ABG) nonsingular black hole}

\author{H.~S.~Ramadhan}
\email{hramad@sci.ui.ac.id}
\author{M.~F.~Ishlah}
\email{muhammad.fauzan79@sci.ui.ac.id.}
\author{F.~P.~Pratama}
\email{fernanda.putra@ui.ac.id.}
\author{I.~Alfredo}
\email{immanuel.alfredo@ui.ac.id.}

\affiliation{Departemen Fisika, FMIPA, Universitas Indonesia, Depok, 16424, Indonesia. }
\def\changenote#1{\footnote{\bf #1}}

\begin{abstract}

We study nonsingular black holes viewed from the point of view of Ayon-Beato-Garcia (ABG) nonlinear electrodynamics (NLED) and present a complete study of their corresponding strong gravitational lensing. The NLED modifies the the photon's geodesic, and our calculations show that such effect increases the corresponding photon sphere radius and image separation, but decreases the magnification. We also show that the ABG's shadow radius is not compatible with bound estimates of Sgr A* from Keck and VLTI (Very Large Telescope Interferometer). Thus, the possibility of Sgr A* being a nonsingular ABG black hole is ruled out. 
\end{abstract}

\maketitle
\thispagestyle{empty}
\setcounter{page}{1}

\section{Introduction}
\label{intro}
Black hole (BH) is one of the most straightforward yet profound prediction of General Relativity (GR). Its extreme gravity distorts its surrounding spacetime and bends light, creating (among many things) the {\it gravitational lensing} phenomenon. The recent observation by Event Horizon Telescope (EHT) that successfully captured the visual images of the superheavy BHs M87*~\cite{EventHorizonTelescope:2019dse} and Sgr A*~\cite{EventHorizonTelescope:2022wkp} has established a triumph for the gravitational lensing as a means to empirically prove black hole's existence. By ``image" here is the corresponding {\it shadow}~\cite{Falcke:1999pj} surrounded by accreting materials that emits and lenses light from the nearby background source. 

Theoretically, the study of gravitational lensing is as old as GR itself (see, for example, \cite{schneider:1992, Perlick:2004tq} and the references therein), but it was Darwin who first applied it for Schwarzschild BH~\cite{Darwin_gravity_1959}. His exact calculation on the deflection angle shows that at small impact parameters there exists a critical value (close to the corresponding photon sphere) where the deflection angle suffers from logarithmic divergence~\cite{Bozza:2010xqn}, beyond which photons fall into the horizon. His results were later rediscovered and developed by other authors, for example in~\cite{Luminet:1979nyg, AbhishekChowdhuri:2023ekr, Ghosh:2022mka, Chandrasekhar:1985kt}. In the last two decades the study of gravitational lensing in the strong deflection limit received revival and extensive elaboration~\cite{Frittelli:1999yf, Bozza:2001xd, Bozza:2002zj, Virbhadra:1999nm}\footnote{Recently, Virbadhra modeled the supermassive black hole $M87^*$ as a Schwarzschild lens and strudied its distorted (tangential, radial, and total) magnifications of images with respect to the change in angular source position and lens-source ratio distance~\cite{Virbhadra:2022iiy}.}. In particular, Bozza shows in~\cite{Bozza:2002zj} that the analytical expansion of the strong deflection angle in the limit of $r\rightarrow r_{ps}$ ($r_{ps}$ being the photon sphere radius) is given by
\begin{equation}
	\alpha(b) = -\tilde{a}_1 \log \left( \frac{b}{b_c} - 1\right) + \tilde{a}_2 + O(b - b_c).
	\label{eq:Defleksi_b}
\end{equation}
with $\tilde{a}_1$ and $\tilde{a}_2$ some constants. Upon closer inspection, Tsukamoto gave correction to the higher order expansion~\cite{Tsukamoto:2016qro, Tsukamoto:2016jzh}. The result on Schwarszchild was extended to the case of Reissner-Nordstrom by Eiroa {\it et al}~\cite{Eiroa:2002mk}, while the strong lensing in Kerr BH was studied in~\cite{Bozza:2002af, Vazquez:2003zm, Bozza:2006nm}.

Probably the most intriguing property of black holes is the existence of singularity due to the gravitational collapse~\cite{Oppenheimer:1939ue}. It was believed that such singularity is an inherent solution of general relativity, but the stable ones (like all observable black holes) are disconnected from the observers by event horizon~\cite{Penrose:1964wq}. Nevertheless, Bardeen in 1968 constructed a metric function that produced nonsingular spacetime~\cite{bardeenNonsingularGeneralrelativisticGravitational1968}. The metric and all invariants are devoid of singularity everywhere, including at $r=0$. Instead, we have regular de Sitter space at the core. (For an excellent review on regular BH see, for example, \cite{Ansoldi:2008jw}.) The strong lensing phenomenon around Bardeen BH has been studied in~\cite{Eiroa:2010wm, Wei:2015qca}. 

At first nobody knows what kind of matter that sources the Bardeen geometry, but later Ayon-Beato and Garcia (ABG) realized that this nonsingular metric can be obtained as solutions of Einstein's equations coupled to some nonlinear electrodynamics (NLED) source~\cite{Ayon-Beato:1998hmi, Ayon-Beato:2000mjt}. Invoking NLED turns out to have profound impact on the geodesic of test photon. Novello {\it et al} showed that in nonlinear electrodynamics background, photon moves in an effective modified geometry~\cite{Novello:1999pg}, and this radically modifies the corresponding optical observables. In~\cite{Allahyari:2019jqz} the authors model the M87* as (singular) NLED-charged black hole and studied its shadow.

In this work, we discuss the effect of effective geometry to the lensing phenomenon in the Bardeen BH using one of the ABG's NLED model. In particular, we calculate the image separation and magnification. We use it as a model for the supermassive black hole at the center of our galaxy, the Sgr A*, and calculate the astrophysical observables. Lastly, we investigate its shadow radius and compare it to the astrophysical data from Keck and VLTI.  This paper is organized as follows. In Sec.~\ref{sec:bardeen} we briefly review the regular Bardeen solution and its corresponding ABG models. In Sec.~\ref{sec:eff} we present the effective metric of ABG and the corresponding photon sphere. Sec.~\ref{sec:deflection} is devoted to applying the Bozza's and Tsukamoto's strong lensing formalism to our model. Sec.~\ref{sec:lensing} is devoted to calculating the strong lensing observables using the Sgr A* data. In Sec.~\ref{sec:shadow} we calculate the shadow radius and plot it against the Keck-VLTI constraints. Finally, we summarize our findings in Sec.~\ref{sec:conc}.


\section{Bardeen Spacetime}
\label{sec:bardeen}
The Bardeen metric is given by~\cite{bardeenNonsingularGeneralrelativisticGravitational1968}:
\begin{equation}
	ds^2=-f(r)dt^2+f(r)^{-1}dr^2+r^2d\Omega^2,
\end{equation}
with
\begin{equation}
	\label{bardeen}
	f(r)\equiv1 - \frac{2mr^2}{(r^2 + q^2)^\frac{3}{2}},
\end{equation}
and $q$ is the charge. This spacetime is regular at $r = 0$, as can easily be seen from the Kretschmann scalar:
\begin{equation}
	\lim_{r\rightarrow0}R^{\alpha\beta\gamma\delta}R_{\alpha\beta\gamma\delta}=\frac{96m^2}{q^{8/3}},    
\end{equation}
while the metric behaves de Sitter-like
\begin{equation}
	f(r)\approx 1-\frac{2m}{q^3}r^2.
\end{equation}
The horizons $r_h$ are given by the roots of
\begin{equation}
	\left(r_h^2+q^2\right)^3-4m^2r_h^4=0.
\end{equation}
Bardeen black hole can, in general, possess up to two horizons. The extremal condition is achieved when~\cite{Ayon-Beato:2000mjt}
\begin{equation}
	q^2=\frac{16}{27}m^2\ \rightarrow r_{extr}=\sqrt{\frac{32}{27}}m.
\end{equation}
Ayon-Beato and Garcia proposed the NLED matter to source the Bardeen spacetime, given in~\cite{Ayon-Beato:1998hmi}. This model, however, produces a slightly different metric function than the original Bardeen,
\begin{equation}
	f(r)=1 - \frac{2mr^2}{(r^2 + q^2)^\frac{3}{2}}+\frac{q^2 r^2}{\left(r+2+q^2\right)^2}.
	\label{ABGori}
\end{equation}
Strong lensing of this particular model has been discussed in~\cite{Ghaffarnejad:2016dlw}. Later ABG considered a simpler NLED sourced by magnetic monopole as follows~\cite{Ayon-Beato:2000mjt},
\begin{equation}
	\mathcal{L} = \frac{3}{ 2 s q^2} \left( \frac{\sqrt{2q^2 F}}{1 + \sqrt{2q^2 F}}\right)^{5/2},
	\label{eq:LagrangeABG}
\end{equation}
where $F\equiv\frac{1}{4}F^{\mu\nu}F_{\mu\nu}$ and $s\equiv q/2m$. By inserting the monopole ansatz $A_{\mu}=\delta^{\varphi}_{\mu}q\left(1-\cos\theta\right)$, the field strength becomes $F=q^2/2r^4$ and the Lagrangian produces the metric solution Eq.~\eqref{bardeen}.
\section{Effective Geometry}
\label{sec:eff}

The NLED Lagrangian above induces the effective metric tensor~\cite{Novello:1999pg} 
\begin{equation}
	g^{\mu\nu}_{eff} = g^{\mu\nu} - \frac{4\mathcal{L}_{FF}}{\mathcal{L}_{F}}F^{\mu}_{\alpha}F^{\alpha\nu},
\end{equation}
where $\mathcal{L}_{A}\equiv\partial\mathcal{L}/\partial A$. This, in turn, yields the effective length element
\begin{equation}
	ds^2_{eff} = -f(r)dt^2 + f(r)^{-1}dr^2 + h_m(r)r^2d^2\Omega,
\end{equation} 
where
\begin{equation}
	h_m(r) = \left(1+\frac{4\mathcal{L}_{FF}}{\mathcal{L}_F}\frac{q^2}{r^4}\right)^{-1}.
\end{equation}
Inserting the Lagrangian~\eqref{eq:LagrangeABG} we obtain
\begin{equation}
	h_{ABG}(r) = \left( 1 - \frac{2(6q^2 - r^2)}{(q^2 + r^2)} \right)^{-1}.
\end{equation}

From the corresponding geodesic equation it is not difficult to see that the radial equation satisfies
\begin{equation}
	\frac{1}{2}\dot{r}^2+V_{eff}=0,
\end{equation}
where we define the effective potential $V_{eff}$ as
\begin{equation}
	V_{eff} \equiv \frac{f(r)}{h(r)}\frac{\mathbb{L}^2}{r^2}.
\end{equation}
The corresponding photon sphere radius is given by the largest positive root of the following condition,~\cite{Bozza:2002zj}
\begin{equation}
	\frac{f'(r_{ps})}{f(r_{ps})} - \frac{2}{r_{ps}} -  \frac{h'(r_{ps})}{h(r_{ps})}=0.
\end{equation}
This yields,
\begin{eqnarray}
	\frac{2}{r_{ps}}+\frac{28q^2r_{ps}}{11q^4+8q^2r_{ps}^2-3r_{ps}^4}+\frac{m\left(4q^2r_{ps}-2r_{ps}^3\right)}{\left(q^2+r_{ps}^2\right)\left[\left(q^2+r_{ps}^2\right)^{3/2}-2mr_{ps}^2\right]}=0.
\end{eqnarray}
\begin{figure}
	\centering
	\includegraphics[scale=0.7]{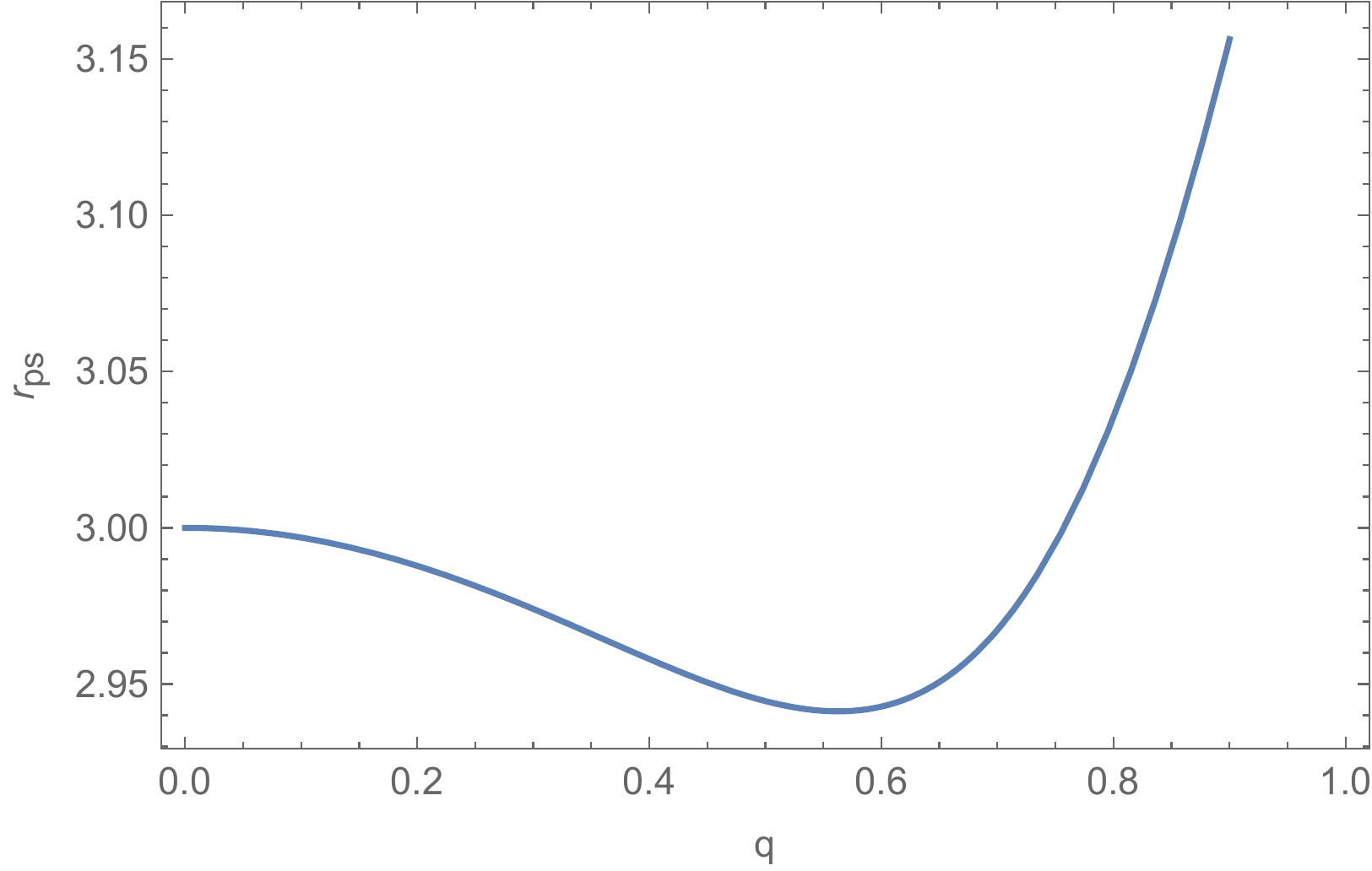}
	\caption{Photon Sphere of Bardeen with ABG source as function of charge ($q$) for $m=1$.}
	\label{fig:ABGphotonsphere1}
\end{figure}
By solving the roots numerically, the behavior of $r_{ps}$ as a function of q,  $r_{ps}=r_{ps}(m=1,q)$, is shown in Fig.\ref{fig:ABGphotonsphere1}. It is shown that the photon sphere decreases as the charge increases until some critical value $q_{crit}$ where $r_{ps}(m=1,q=q_{crit})$ is minimum, beyond which $r_{ps}$ starts increasing without bound. Interestingly, the critical value $q_{crit}$ is not given by the extremal charge $q=\sqrt{16/27}m$, as in the Bardeen case. Rather, $q_{crit}=16/27\ m=0.592$. \begin{figure}
	\centering
	\includegraphics[scale=0.8]{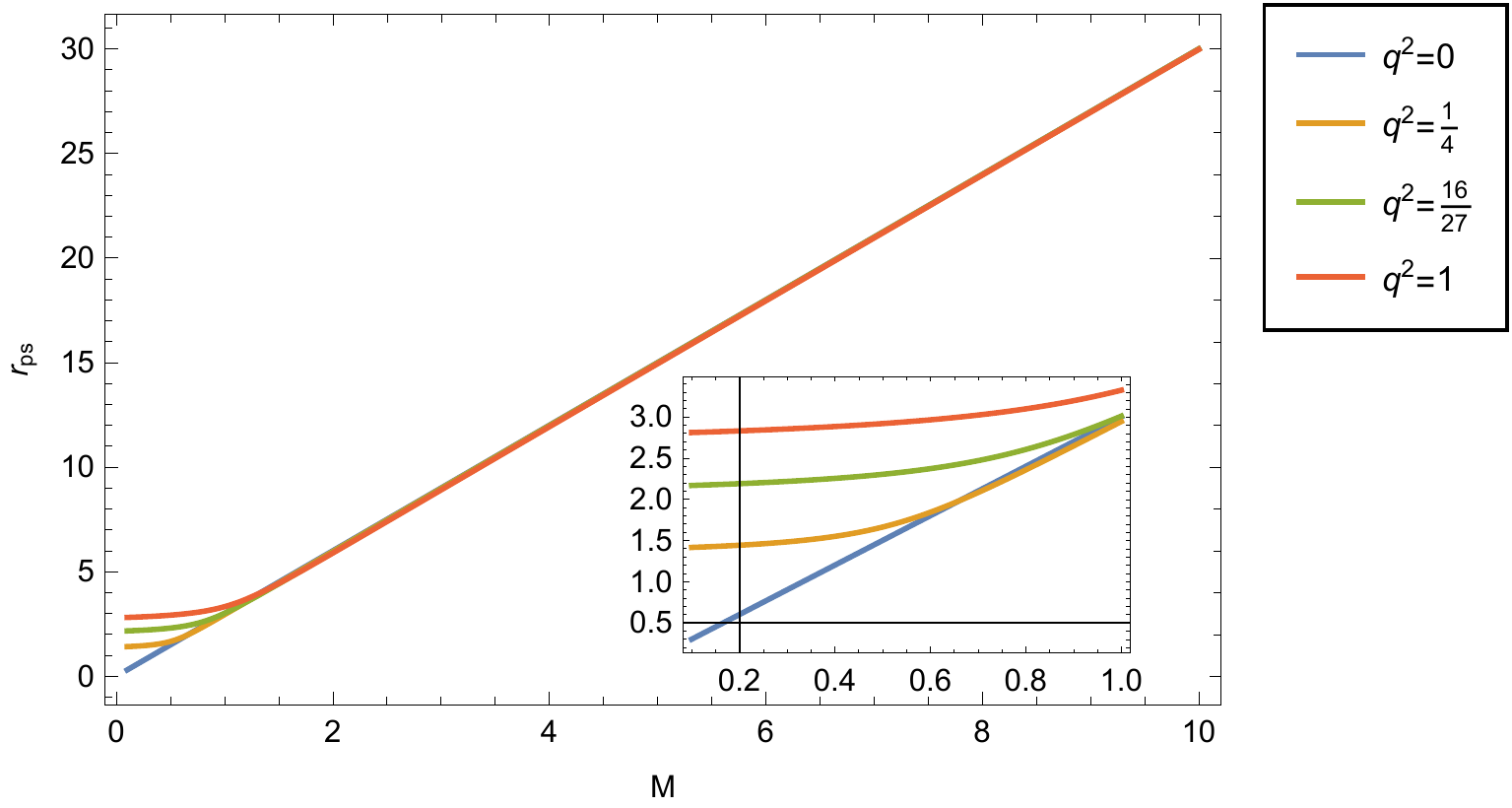}
	\caption{$r_{ps}$ as function of $m$ for a variation of charge $q$.}
	\label{fig:grafik_rpsm}
\end{figure}
\begin{figure}
	\centering
	\includegraphics[scale=1.0]{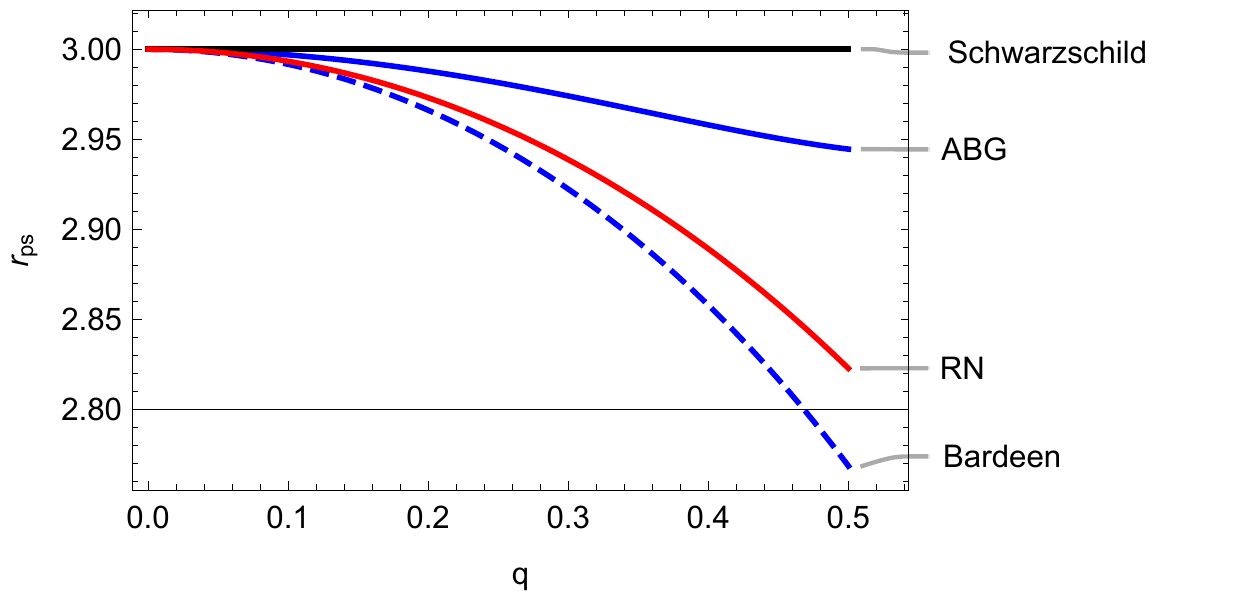}
	\caption{$r_{ps}$ as function of $q$ for different models.}
	\label{fig:grafik_rpsfull}
\end{figure}
In Fig.~\ref{fig:grafik_rpsm} we show how $r_{ps}$ varies with $m$ for several values of $q$. They differ only at small $m$. When the mass is large, $r_{ps}$ for different $q$ asymptote to a single gradient. In Fig.~\ref{fig:grafik_rpsfull} we show the deviation of $r_{ps}$ as a function of $q$ from the Schwarzschild (charge-less condition). The ABG model we consider here falls between the Schwarzschild and the RN at large $q$, unlike the original Bardeen which falls the fastest.
\section{Deflection Angle in the Strong Field Limit}
\label{sec:deflection}

From the spherical symmetry and staticity conditions, Noether's theorem dictates that this spacetime has constants of motion, the total test particle's energy $E$ and angular momentum $\mathbb{L}$, related to the $\partial_t$ and $\partial_{\varphi}$ Killing vectors, respectively. We define the impact parameter for photon as 
\begin{equation}
	\label{critimpact}
	b(r_0)\equiv\frac{\mathbb{L}}{E}=\sqrt{\frac{h_m(r_0)r_0^2}{f(r_0)}}.
\end{equation}

Solving the null geodesic equation, the general expression for  bending angle of light rays can be expressed as (see, for example, in~\cite{Weinberg:1972kfs})
\begin{equation}
	\label{alpha}
	\alpha(r_0) = 2 \int_{r_0}^\infty \sqrt{\frac{1}{r^2 f(r) h(r) R(r)}}dr - \pi, 
\end{equation}
where
\begin{equation}
	R(r) = \frac{r^2 h(r)}{b^2 f(r) } - 1. 
\end{equation}
The integral is divergent at $r_0\rightarrow r_{ps}$. To circumvent this problem we define $z\equiv1-r_0/r$~\cite{Tsukamoto:2016jzh} 
and write Eq.~\eqref{alpha} as
\begin{equation}
	\label{ar0}
	\alpha(r_0) = \int^1_0\mathcal{H}(z,r_0) dz-\pi,
\end{equation}
with 
\begin{equation}
	\mathcal{H}(z,r_0) \equiv  \frac{2\left(1-z\right)^2}{\sqrt{f\left(\frac{r_0}{1-z}\right)h\left(\frac{r_0}{1-z}\right)R\left(\frac{r_0}{1-z}\right)}}.
\end{equation}
The singular part can be isolated by defining
\begin{equation}
	\mathcal{H}(z,r_0)\equiv\mathcal{H}_R(z,r_0)+\mathcal{H}_D(z,r_0),
\end{equation}
where the subscript R(D) refers to the regular (divergent) part, respectively.

To handle the divergent part: 
\begin{equation}
	\mathcal{I}_D(r_0)\equiv\int^1_0\mathcal{H}_D(z,r_0)dz,
\end{equation}
we define $\mathcal{H}(z,r_0)\equiv2r_0/\sqrt{\mathcal{G}(z,r_)}$ and, by expanding it around $z\rightarrow0$, obtain the expression for $\mathcal{G}(z,r_0)$ up to second-order:
\begin{equation}
	\label{g0}
	\mathcal{G}_0(z,r_0)=c_1(r_0)z+c_2(r_0) z^2,
\end{equation}
where
\begin{eqnarray}
	c_1(r_0)&=& C_0\mathcal{D}_0r_0 f(r_0),\nonumber\\
	c_2(r_0)&=&C_0r_0f_0\bigg\{\mathcal{D}_0\left[\mathcal{D}_0\left(\mathcal{D}_0+\frac{f'_0}{f_0^3}\right)r_0-3\right]+\frac{\mathcal{D}_0r_0}{2}\bigg\},
\end{eqnarray}
with $X_0\equiv X(r=r_0)$, $C(r)\equiv h(r)r^2$, and
\begin{equation}
	\mathcal{D}(r)\equiv\frac{C''(r)}{C(r)}-\frac{f''(r)}{f(r)}.
\end{equation}
For the ABG model, the values of $c_1$ and $c_2$ are
\begin{widetext}
	\begin{eqnarray}
		c_1(r_0) &=& - \frac{2(r_0^2 (11 q^4 + 22 q^2 r_0^2 - 3r_0^4) \tilde{j}_1^{3/2}+ m(9r_0^8 - 61 q^2 r_0^6)))}{(11 q^2 - 3r^2)^2\tilde{j}_1^{3/2}},\nonumber\\
		c_2(r_0) &=& \frac{m r_0^6\tilde{j}_2 + r_0^2(\tilde{j}_3 \tilde{j}_1^{5/2} - \tilde{j}_3)}{(11q^2 - 3r_0^2)^3 \tilde{j}_1^{5/2}},\nonumber\\
	\end{eqnarray}
\end{widetext}
with
\begin{eqnarray}
	\tilde{j}_1 &\equiv& q^2 + r_0^2 ,\nonumber\\
	\tilde{j}_2 &\equiv& 2013 q^6 + 3104 q^4 r_0^2 + 57 q^2 r_0^4 - 54 r_0^6,\nonumber\\
	\tilde{j}_3 &\equiv& 121q^6 + 825 q^4 r_0^2 - 99 Q^2 r_0^4 + 9 r_0^6.
\end{eqnarray}
Following Bozza~\cite{Bozza:2002zj} and Tsukamoto~\cite{Tsukamoto:2016jzh} it can be shown that the divergent integral in the strong field limit $r\rightarrow r_{ps}$ (or equivalently $b\rightarrow b_{ps}$) is expressed as
\begin{eqnarray}
	\label{id}
	\mathcal{I}_D(b)&=&-\sqrt{\frac{2}{f_{ps}C_{ps}''-f''_{ps}C_{ps}}}\nonumber\\
	&&\times\bigg\{\log\left(\frac{b}{b_c}-1\right)+\log\left[r_{ps}^2\left(\frac{C_{ps}''}{C_{ps}}-\frac{f_{ps}''}{f_{ps}}\right)\right]\bigg\}.\nonumber\\
\end{eqnarray}
\begin{figure}
	\centering
	\includegraphics[scale=0.8]{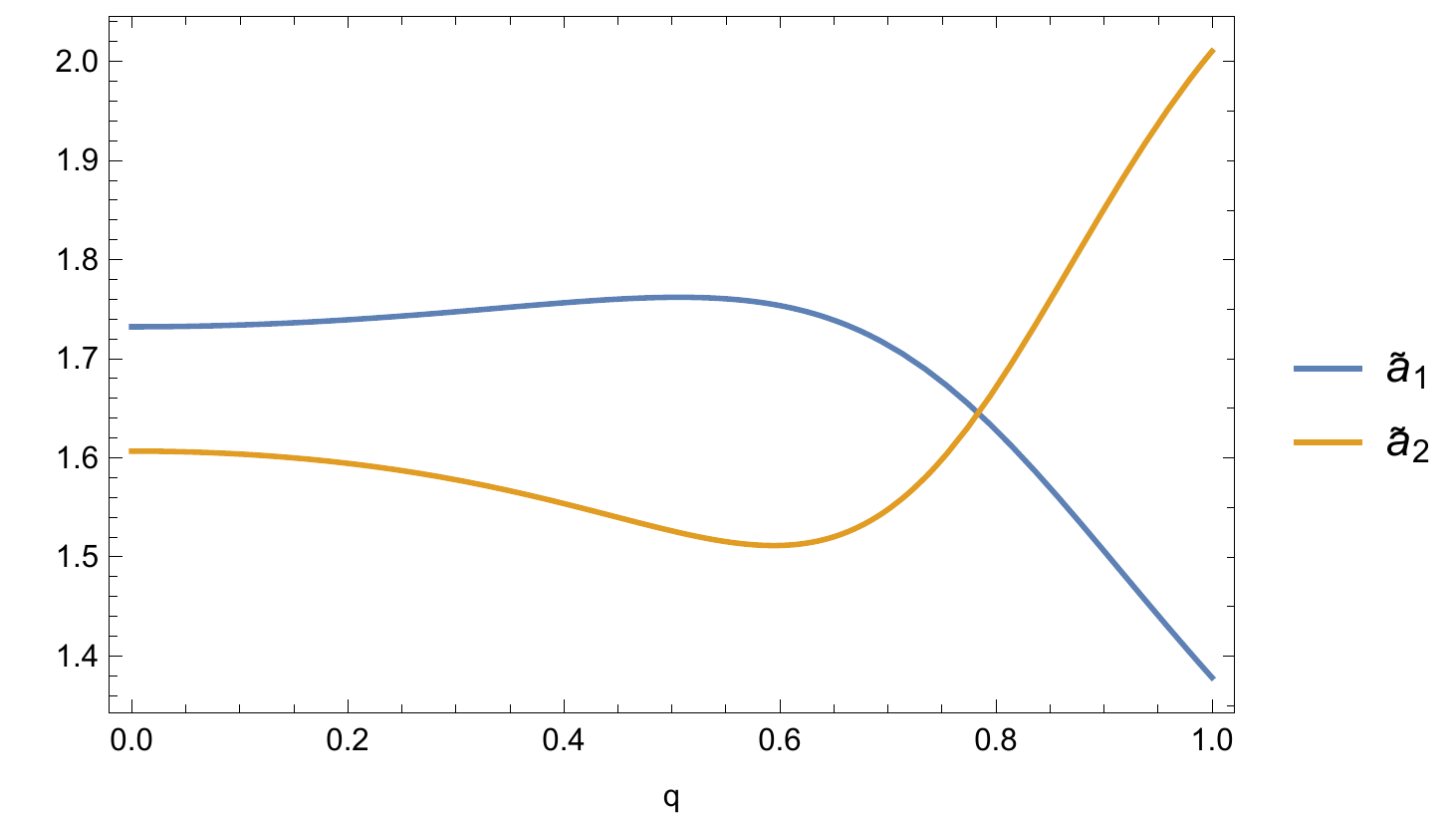}
	\caption{The values of $\tilde{a}_1$ and $\tilde{a}_2$ for Bardeen with ABG source}
	\label{fig:a1a2ABG}
\end{figure}

The regular part is
\begin{equation}
	\mathcal{I}_R\equiv\int^1_0\mathcal{H}_R(z,r_0)dz.
\end{equation}
Likewise, by expanding around $r\rightarrow r_{ps}$ the integral can be expressed as
\begin{eqnarray}
	\label{ir}
	\frac{\mathcal{I}_R(r_{ps})}{2r_{ps}}&=&\int^1_0\frac{dz}{F(r_{ps},z)}-\int^1_0\sqrt{\frac{2}{C_{ps}r_{ps}f_{ps}\mathcal{D}_{ps}z^2}}dz,\nonumber\\
\end{eqnarray}
with
\begin{equation}
	F(r_{ps},z)\equiv\sqrt{C\left(\frac{r_{ps}}{1-z}\right)R\left(\frac{r_{ps}}{1-z}\right)f\left(\frac{r_{ps}}{1-z}\right)\left(1-z\right)^4}.
\end{equation}
For the ABG model, the integral can be evaluated numerically. Putting~\eqref{id} and~\eqref{ir} into~\eqref{ar0}, we can express it as Eq.~\eqref{eq:Defleksi_b} by identifying
\begin{eqnarray}
	\tilde{a}_1&\equiv&\sqrt{\frac{2 A_{ps} B_{ps}}{A_{ps}C''_{ps} - A''_{ps} C_{ps}}},\nonumber\\
	\tilde{a}_2&\equiv&\tilde{a_1} \log \tilde{b} + I_R(r_{ps}) - \pi,
\end{eqnarray}
with
\begin{equation}
	\tilde{b}\equiv r_{ps}^2 \left( \frac{C''_{ps}}{C_{ps}} -\frac{ A''_{ps}}{A_{ps}} \right).
\end{equation}
In terms of the ABG model we consider, their expressions are
\begin{eqnarray}
	\tilde{a}_1&=&\sqrt{\frac{(11q^2 - 3r_{ps}^2)^3 \tilde{k}_1^{5/2}}{\tilde{k}_2 - \tilde{k}_3 \tilde{k}_1^{5/2}}},\nonumber\\
	\tilde{a}_2&=&- \pi+ \mathcal{I}_R(r_{ps})+\sqrt{\frac{(11q^2 - 3r_{ps}^2)^3 \tilde{k}_1^{5/2}}{\tilde{k}_2 - \tilde{k}_3 \tilde{k}_1^{5/2}}}\log\left[\frac{2 \tilde{k}_1^{5/2} \tilde{k}_3 - 2 \tilde{k}_2}{(11 q^2 - 3 r_{ps}^2)^2 (\tilde{k}_1^{3/2} - 2 m r_{ps}^2) \tilde{k}_1^2}\right],\nonumber\\
\end{eqnarray}
with
\begin{eqnarray}
	\tilde{k}_1 &=& q^2 + r_{ps}^2,\nonumber\\
	\tilde{k}_2 &=& m(3355 q^6 r_{ps}^4 + 466 q^4 r_{ps}^6 + 51 q^2 r_{ps}^8), \nonumber\\
	\tilde{k}_3 &=& (121 q^6 + 825 q^4 r_{ps}^2 - 99 q^2 r_{ps}^4 + 9 r_{ps}^6).
\end{eqnarray}
\begin{figure}
	\centering
	\includegraphics[scale=0.6]{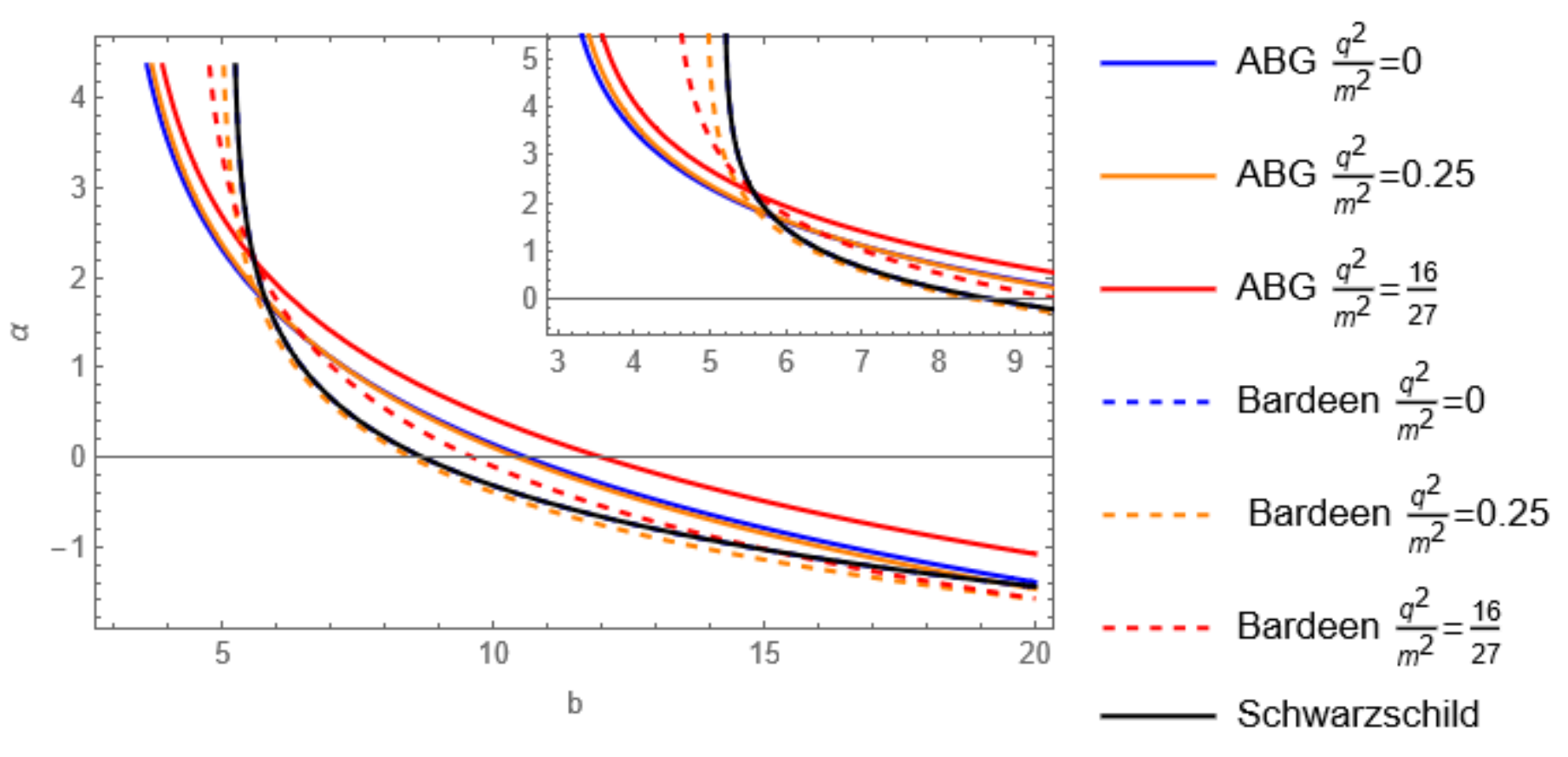}
	\caption{Comparison of light deflection for Regular Bardeen and Bardeen with ABG source}
	\label{fig:LightDeflectionComparison}
\end{figure}

The calculated $\tilde{a}_1$ and $\tilde{a}_2$ are shown in Fig.~\ref{fig:a1a2ABG}. The $\tilde{a}_1$ slowly increases until $q=q_{crit}$ and then decreases linearly. The $\tilde{a}_2$, on the other hand, decreases as $q$ goes up until $q=q_{crit}$, then it starts increasing. The deflection angles are depicted in Fig.~\ref{fig:LightDeflectionComparison}. It is shown that the critical impact parameters for the NLED models are smaller than for the pure Bardeen. This critical value increases with increasing charge. 

\section{Lensing of Sgr A* as an ABG Black Hole}
\label{sec:lensing}

\begin{figure}
	\centering
	\includegraphics[scale=0.7]{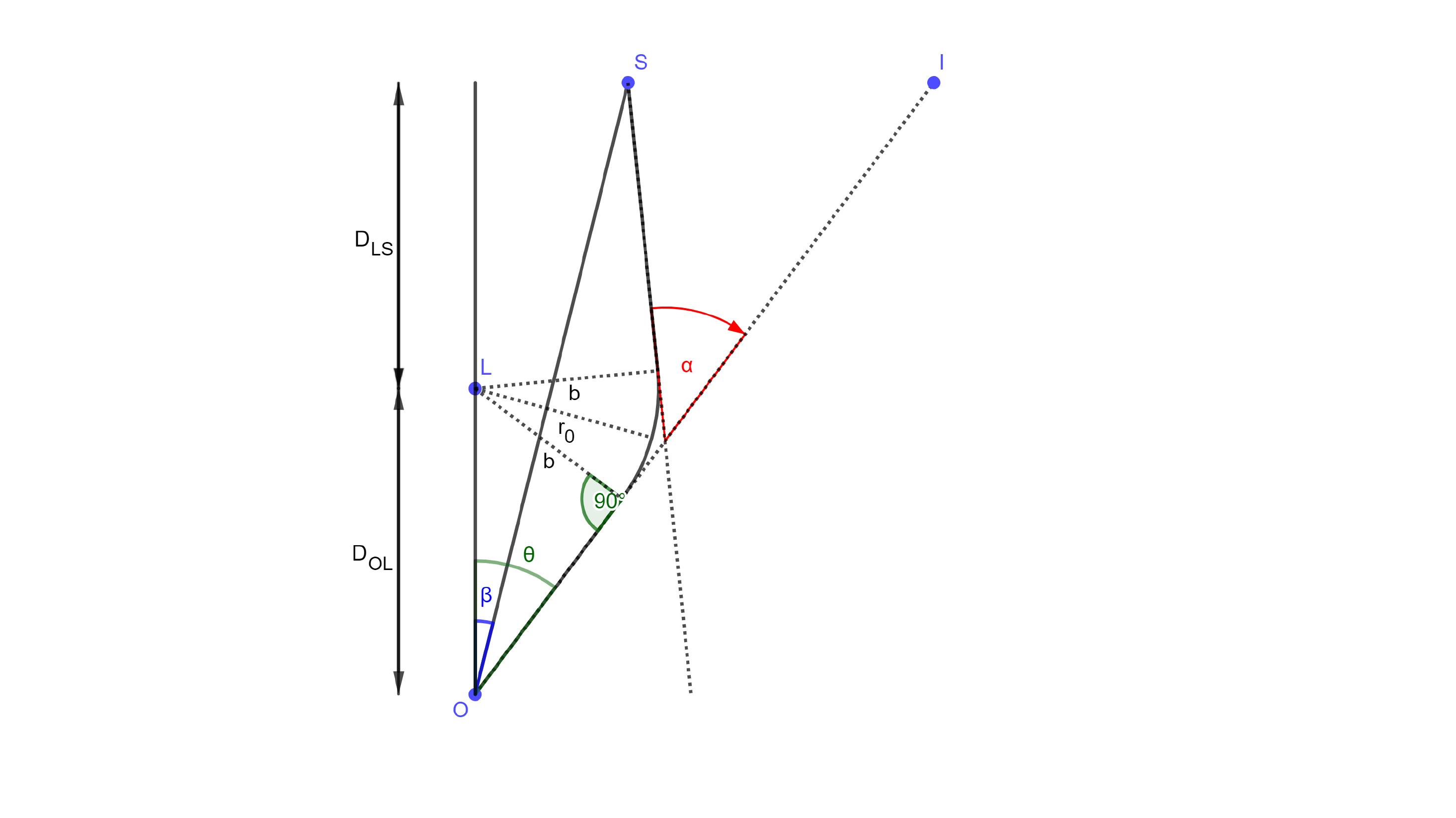}
	\caption{Gravitational lensing diagram. S, L, and O are the source, the lens, and the observer, respectively.}
	\label{fig:light_deflection}
\end{figure}

The most straightforward effect of light deflection due to gravitational field is the notion of ``gravitational lensing". The lensing mechanism can be inferred from Fig.~\ref{fig:light_deflection}. The straight segment $\overline{SO}$ is the path the light would have taken had it not been deflected due to the lens (BH) at $L$. The angle $\beta$ denotes the angular position of the source $S$ from the observer $O$ if there were no lensing. What the $O$ observes is the ``image" of $S$ located at $I$ whose angular position is given by $\theta$. The deflection angle is given by $\alpha$. From simple geometry the relation between $\beta$ and $\theta$ can be expressed as~\cite{Virbhadra:1999nm, Bozza:2008ev}
\begin{equation}
	\tan \beta = \tan \theta - \frac{D_{LS}}{D_{OL}} \left[ \tan \theta + \tan (\alpha - \theta)\right],
\end{equation}
known as the lens equation.

In the strong field limit ($r_0\rightarrow r_{ps}$) $\beta$ and $\theta$ are small, $\alpha $ can exceed $2\pi$ and light can loop around the black hole several ($n$) times before escaping out to the observer. 
In this sense, $\alpha = 2n \pi + \Delta \alpha_n$. We can then expand $\tan (\alpha - \theta)\sim\Delta\alpha_n-\theta$~\cite{Bozza:2001xd}. The lens equation thus becomes
\begin{equation}
	\beta = \theta - \frac{D_{LS}}{D_{OL}} \Delta \alpha_n.
	\label{eq:Persamaan_Lensa}
\end{equation}
We also have the relation
\begin{equation}
	b = D_{OL} \theta.
	\label{eq:impactparameter_posisi}
\end{equation}
Substituting it into \eqref{eq:Defleksi_b} and inverting it results in~\cite{Bozza:2002zj}
\begin{equation}
	\theta(\alpha) \simeq \frac{b_c}{D_{OL}} \left(1 + e^{\frac{\tilde{a}_2 - \alpha}{\tilde{a}_1}}\right).
	\label{eq:Theta_defleksi}
\end{equation}

From Fig.~\ref{fig:light_deflection}, the innermost image is given by
\begin{equation}
	\theta_\infty = \frac{b_c}{D_{OL}}.
	\label{eq:theta_infty}
\end{equation}
Expanding around $\alpha$ yields
\begin{equation}
	\theta_n = \theta_n^0 - \gamma_n \Delta\alpha_n,
	\label{eq:Posisi_bayangan}
\end{equation}
with
\begin{eqnarray}
	\theta_n^0 &=& \frac{b_c}{D_{OL}} \left( 1 + e_n\right),\\
	\gamma_n &=& \frac{b_c}{\tilde{a}_2 D_{OL}}e_n,\\
	e_n &=& e^{\frac{\tilde{a}_2 - 2n \pi}{\tilde{a}_1}}.
\end{eqnarray}
We can eliminate $\Delta\alpha_n$ using the Eq.~\eqref{eq:Persamaan_Lensa}. This results in the equation for the n-th shadow position~\cite{Bozza:2001xd, Bozza:2002zj, Tsukamoto:2016qro}
\begin{equation}
	\theta_n = \theta_n^0 + \frac{ D_{LS}}{D_{OS}}\gamma_n(\beta - \theta_n^0),
\end{equation}
where the second term on the right-hand side is small compared to the first one. For the Einstein ring case, $\beta=0$ and
\begin{equation}
	\theta_{nE} = \left(1- \frac{D_{LS}}{D_{OS}} \gamma_n \right) \theta_n^0.
\end{equation}

The observables we wish to calculate are the image separation and the magnification. The separation $s$ is the difference between the outermost and innermost images,
\begin{equation}
	s = \theta_1 -\theta_\infty = \theta_\infty e^{\frac{\tilde{a}_2 - 2 \pi}{\tilde{a}_1}}.
	\label{eq:O_s}
\end{equation}
The magnification $\mu$ is the {\it inverse} of the corresponding Jacobian determinant for the critical curve~\cite{schneider:1992,Bozza:2001xd}. The $n^{th}$ image magnification is defined to be
\begin{equation}
	\mu_n = \frac{1}{|\det J_{\theta_n^0}|} = \frac{1}{\frac{\beta}{\theta_n^0} \left. \frac{\partial \beta}{\partial \theta}\right|_{\theta_n^0}},
\end{equation}
from which the (relativistic) flux ratio is expressed as
\begin{equation}
	r = \frac{\mu_1}{\sum\limits_{n=2}^\infty \mu_n} = e^{\frac{2\pi}{\tilde{a}_1}}.
	\label{eq:O_RasioFluks}
\end{equation}

In this paper we calculate the strong lensing from the Sgr A* black hole modeled as the Bardeen with ABG source. We use data from GRAVITY collaboration where the black hole mass and its distance from the Earth (observer) are $m=4.154\times 10^6 M_\odot$ and $D_{OL} = 8.178\  kpc$, respectively~\cite{abuterGeometricDistanceMeasurement2019}. These values are consistent with the EHT results~\cite{EventHorizonTelescope:2022wkp}. In Table~\ref{tabel:data_contohlensa} we show the observables. Here the value the magnification is converted to magnitudes $r_m = 2.5 \log_{10} r$~\cite{Bozza:2002zj}. From the Table it can be seen that the observable values for the Bardeen does not differ much from that of RN. However, the observables for the ABG are significantly different from both Bardeen and RN; {\it i.e.,} the ABG's are smaller. This shows that the NLED effect is quite significant here. 
\begin{table}
	\centering
	\begin{tabular}{|c|c|c|c|c|}
		\hline
		{Model} & $Q/m$ & $\theta_\infty$ ($\mu$ as) & $s$($\mu$ as) & $r_m$
		\\ \hline
		{Schwarzschild} & - & 26.0592 & 0.0327 & 6.8184
		\\ \hline 
		\multicolumn{1}{|c|}{{Bardeen}} &0    & 26.0592 & 0.0327 & 6.8184
		\\ \cline{2-5} 
		\multicolumn{1}{|c|}{} & 0.1 & 26.0156 & 0.0332351 & 6.79937
		\\ \cline{2-5} 
		\multicolumn{1}{|c|}{} & 0.2 & 25.8833 & 0.0349143 & 6.74083
		\\ \cline{2-5} 
		\multicolumn{1}{|c|}{} & 0.3 & 25.6569 & 0.0381989 & 6.63828
		\\ \cline{2-5} 
		\multicolumn{1}{|c|}{} & 0.4 & 25.3266 & 0.044135 & 6.48266
		\\ \cline{2-5} 
		\multicolumn{1}{|c|}{} & 0.5 & 24.8751 & 0.0552564 & 6.25695
		\\ \cline{2-5} 
		\multicolumn{1}{|c|}{} & 0.6 & 24.2718 & 0.0788401 & 5.92697
		\\ \cline{2-5} 
		\multicolumn{1}{|c|}{} & 0.7 & 23.4566 & 0.144124 & 5.41046
		\\ \hline
		\multicolumn{1}{|c|}{{RN}} & 0 & 26.0592 & 0.0327 & 6.8184
		\\ \cline{2-5} 
		\multicolumn{1}{|c|}{} & 0.1 & 26.0157 & 0.0330446 & 6.81081
		\\ \cline{2-5} 
		\multicolumn{1}{|c|}{} & 0.2 & 25.8841 & 0.0340718 & 6.7875
		\\ \cline{2-5} 
		\multicolumn{1}{|c|}{} & 0.3 & 25.6612 & 0.0359467 & 6.74699
		\\ \cline{2-5} 
		\multicolumn{1}{|c|}{} & 0.4 & 25.3412 & 0.0389696 & 6.68646
		\\ \cline{2-5} 
		\multicolumn{1}{|c|}{} & 0.5 & 24.9146 & 0.043696 & 6.60104
		\\ \cline{2-5} 
		\multicolumn{1}{|c|}{} & 0.6 & 24.3668 & 0.0511167 & 6.48246
		\\ \cline{2-5} 
		\multicolumn{1}{|c|}{} & 0.7 & 23.6747 & 0.0628463 & 6.31574
		\\ \hline
		{ABG} & 0 & 15.0453 & 1.01295 & 3.93662
		\\ \cline{2-5} 
		& 0.1 & 15.0593 & 1.015 & 3.93252
		\\ \cline{2-5} 
		& 0.2 & 15.1029 & 1.02079 & 3.92061
		\\ \cline{2-5} 
		& 0.3 & 15.1806 & 1.02911 & 3.90248
		\\ \cline{2-5}
		& 0.4 & 15.3007 & 1.03751 & 3.88228
		\\ \cline{2-5} 
		& 0.5 & 15.4768 & 1.04185 & 3.87016
		\\ \cline{2-5} 
		& 0.6 & 15.7286 & 1.03655 & 3.8886
		\\ \cline{2-5} 
		& 0.7 & 16.0813 & 1.01629 & 3.9788
		\\ \hline
	\end{tabular}
	\caption{Observables for the Sgr A* Schwarzschild, Bardeen, RN, and ABG. The $\theta$ and its corresponding separation are expressed in $\mu$ arc-second (as) unit.}
	\label{tabel:data_contohlensa}
\end{table}

\begin{figure}
	\centering
		\includegraphics[scale=0.6]{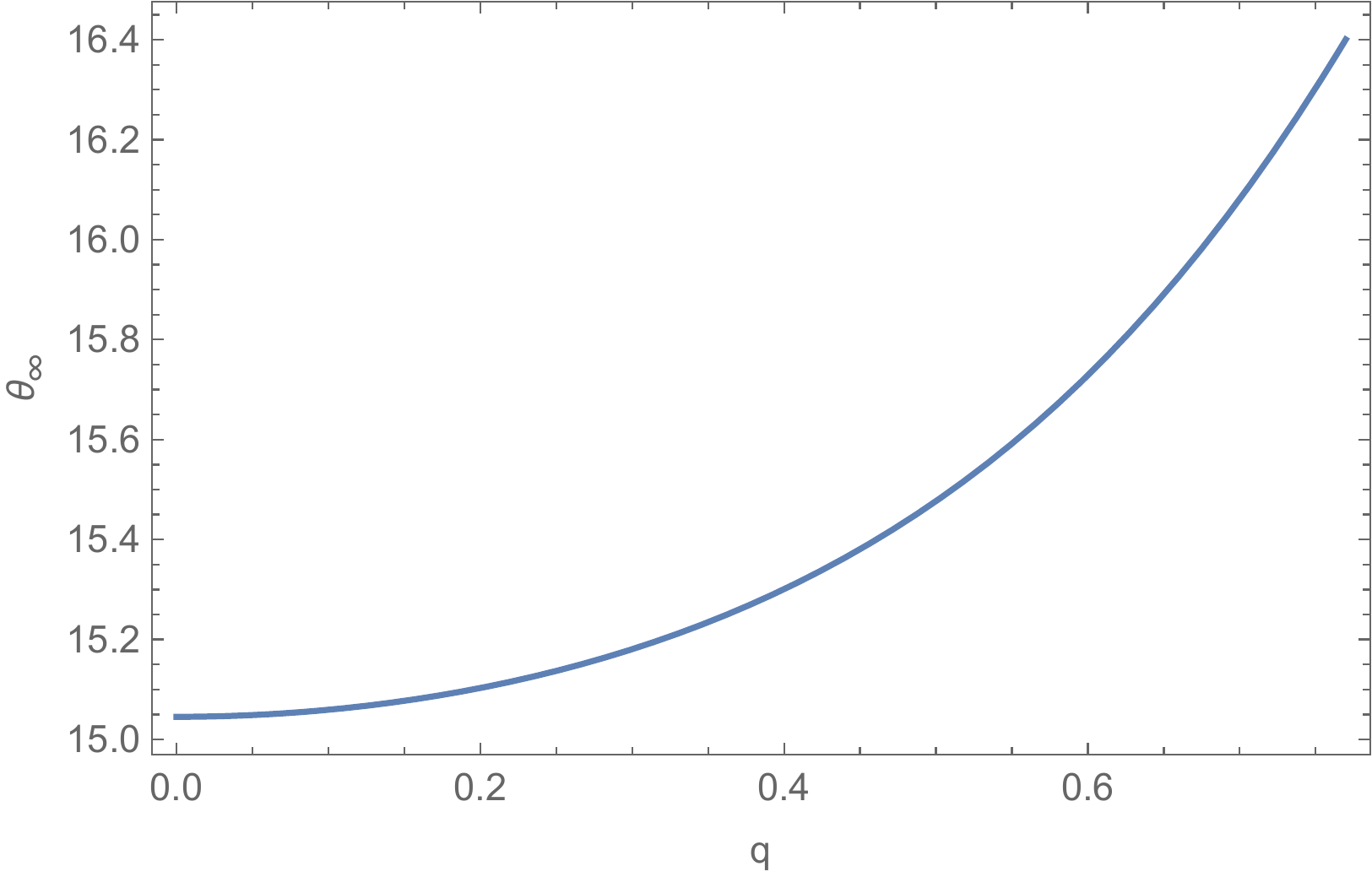} \\
		\includegraphics[scale=0.6]{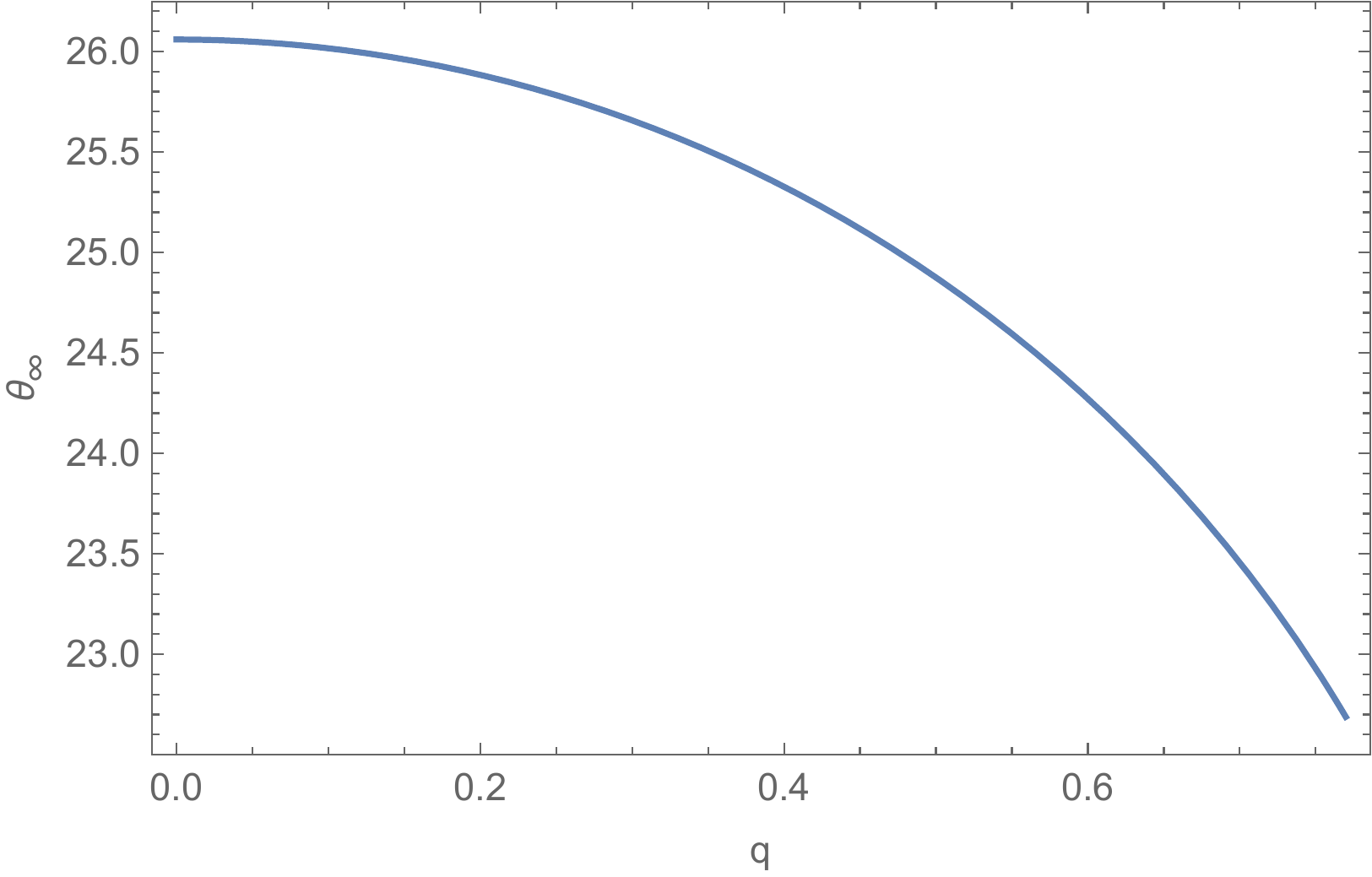}
	\caption{$\theta_\infty$ as function of $q$ for: [top] The ABG model and, [down] the original Bardeen.}
	\label{fig:thetainf}
\end{figure}

\begin{figure}
	\centering
		\includegraphics[scale=0.6]{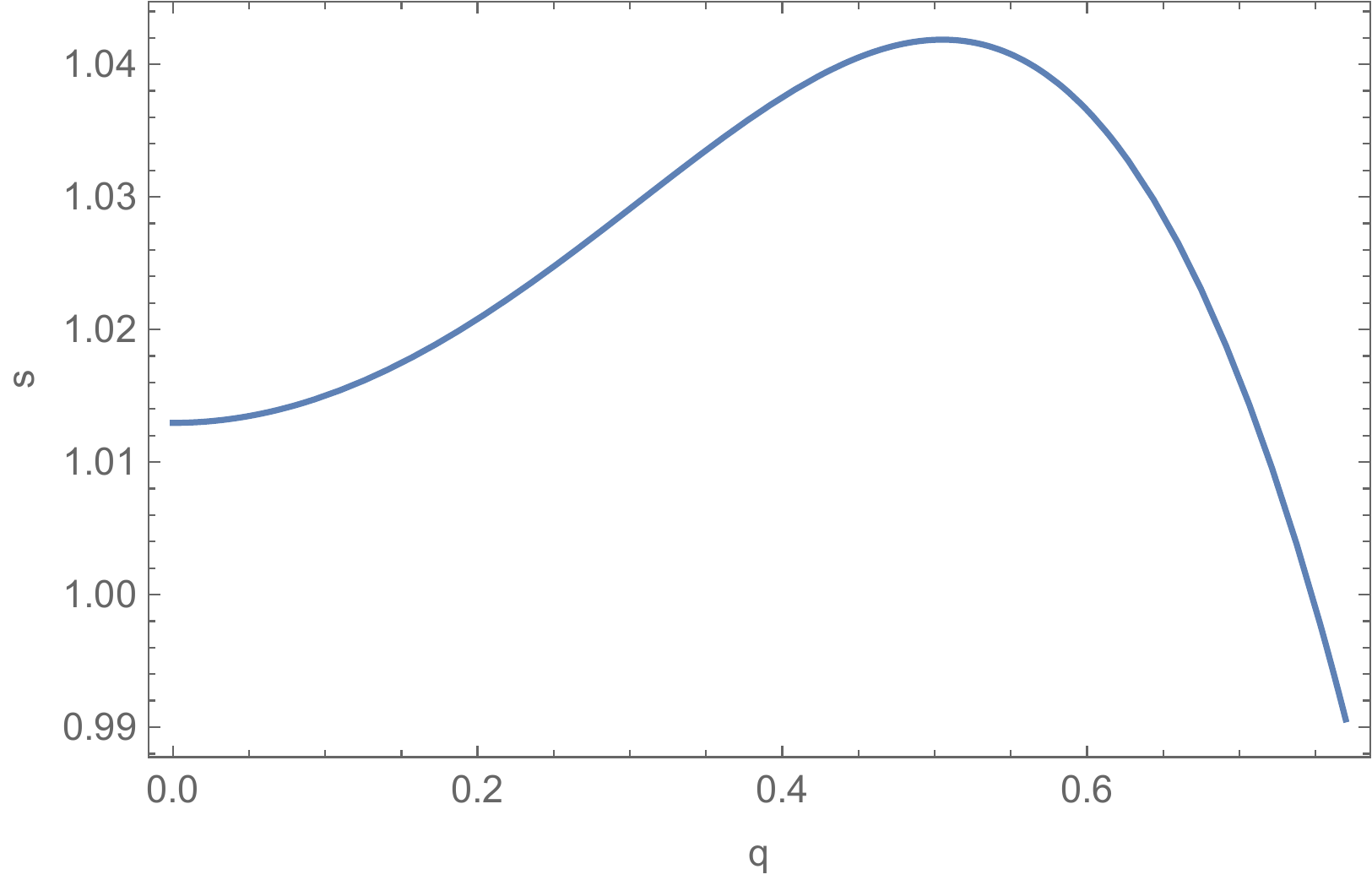} \\
		\includegraphics[scale=0.6]{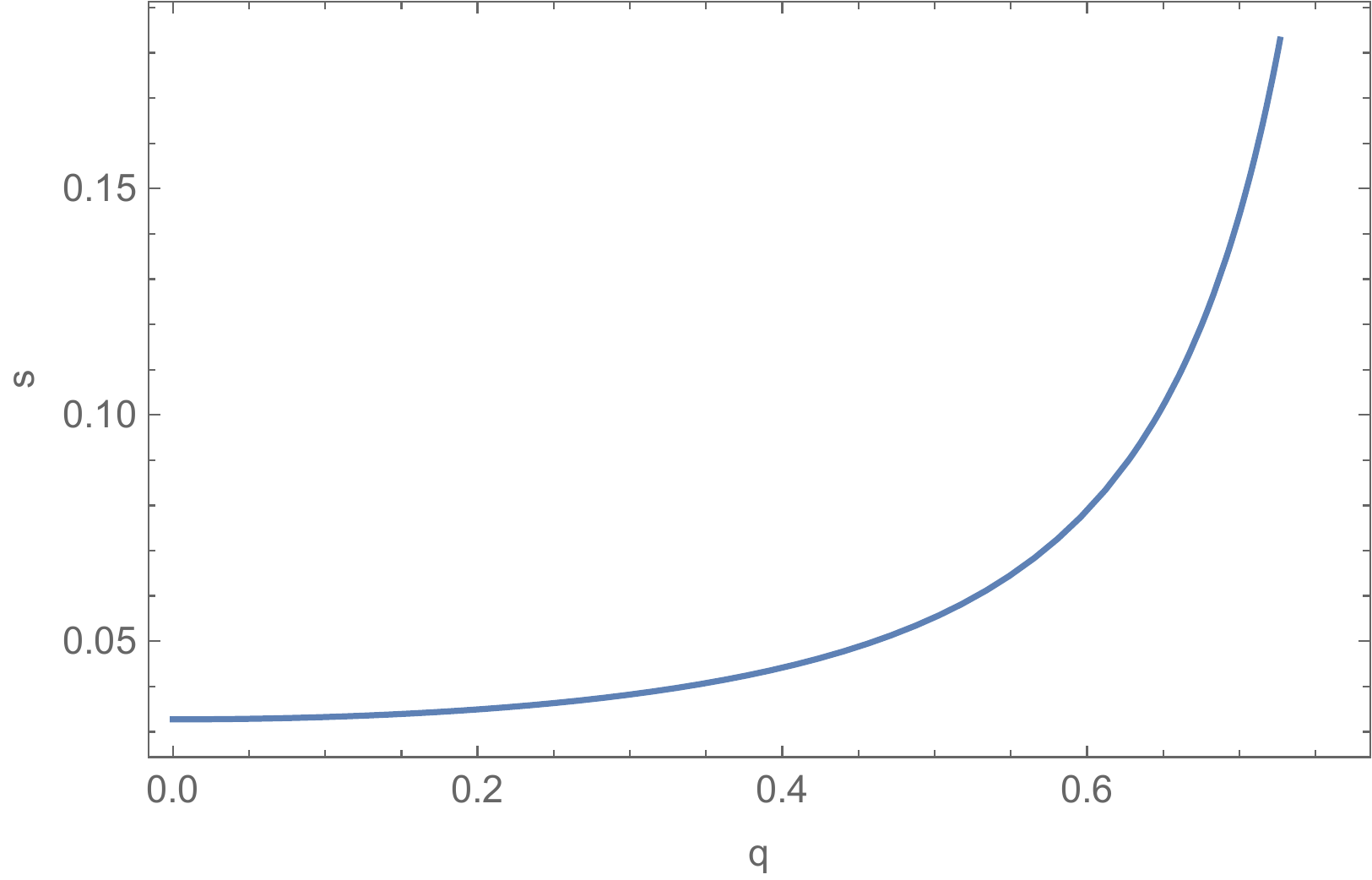}
	\caption{$s$ as function of $q$ for: [top] the ABG model, [down] the Bardeen model.}
	\label{fig:swithq}
\end{figure}

\begin{figure}
	\centering
		\includegraphics[scale=0.6]{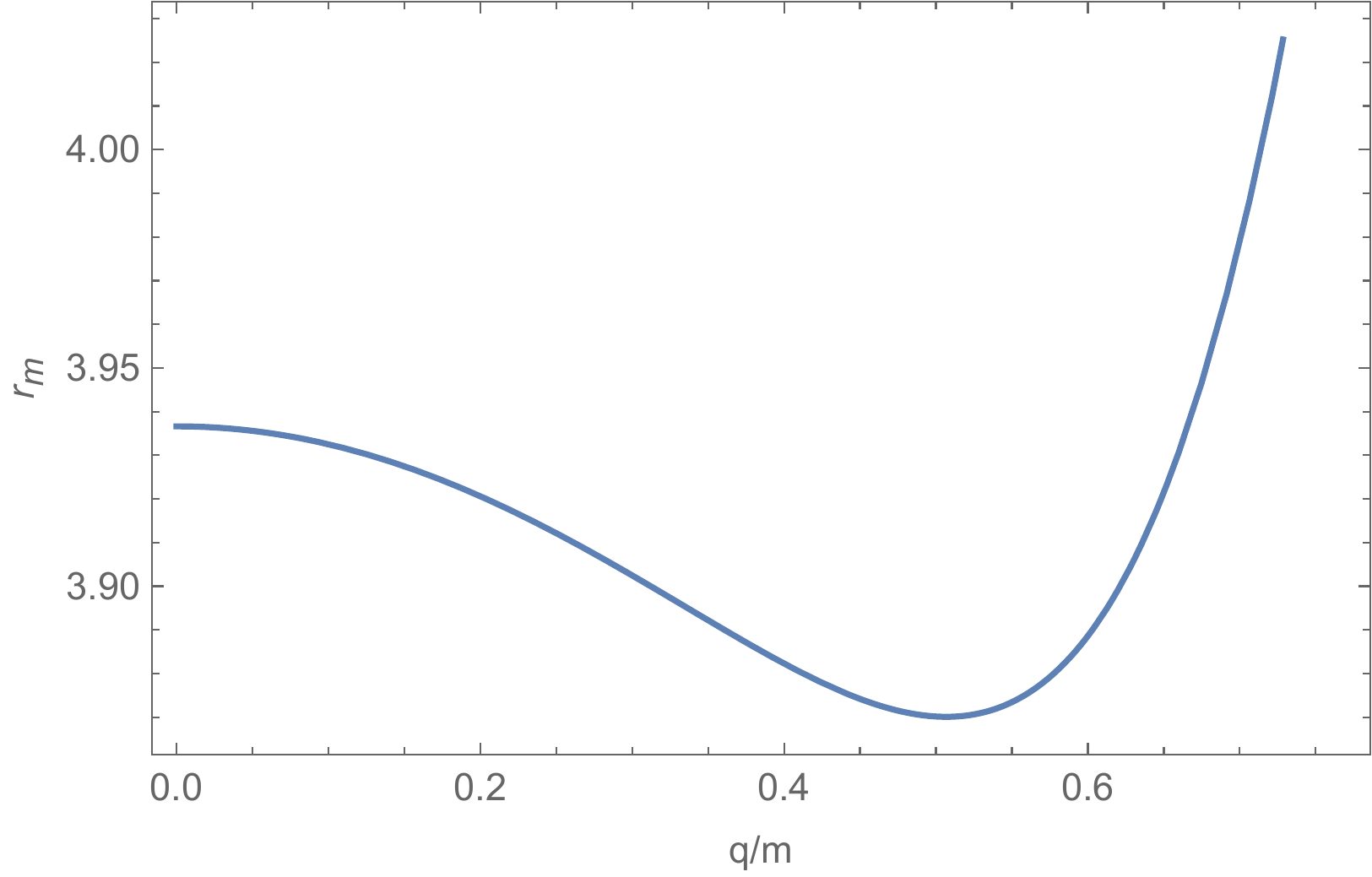} 
		\includegraphics[scale=0.6]{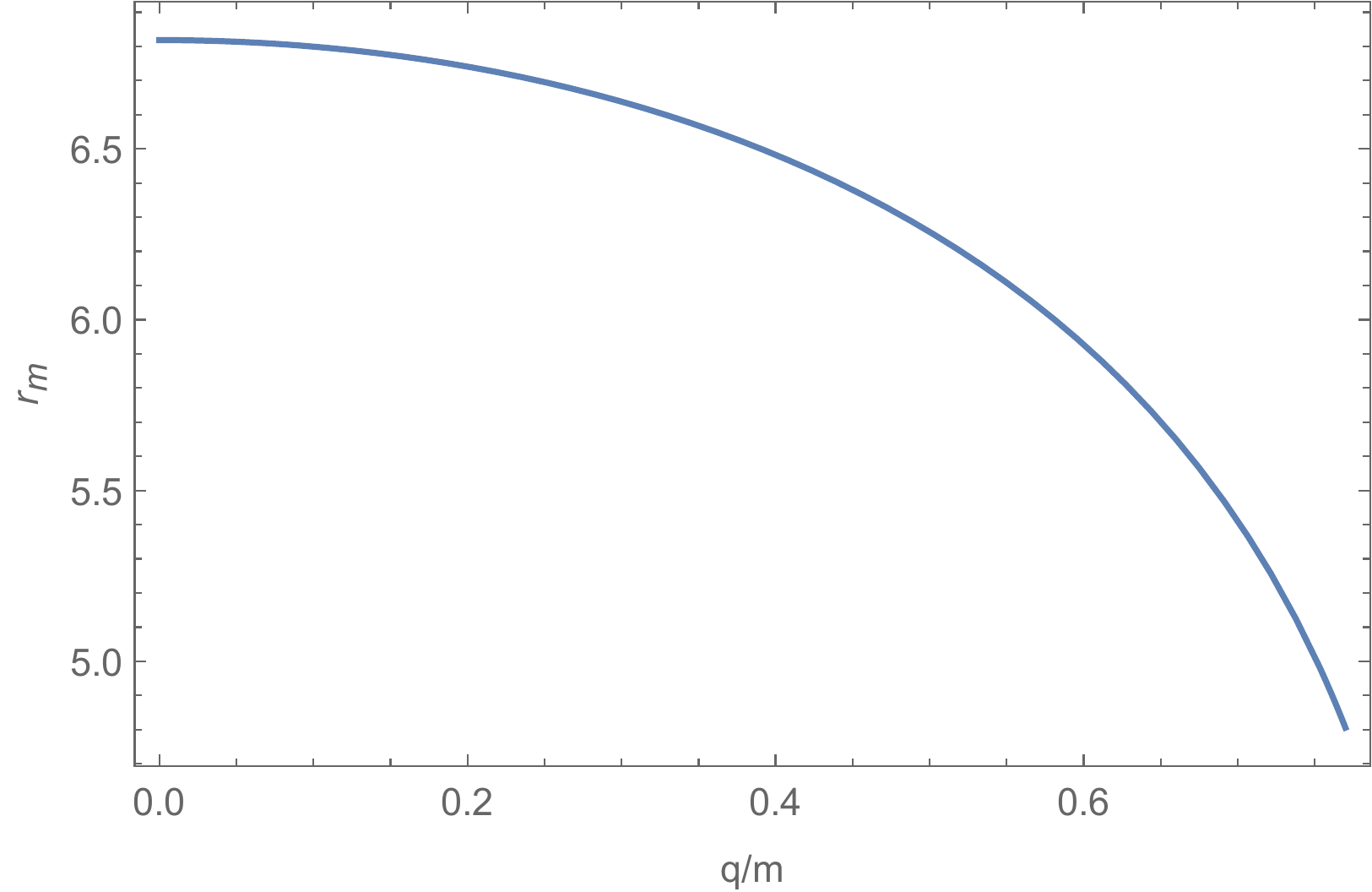}
	\caption{$r_m$ as function of $q$ for: [top] the ABG model, [down] the Bardeen model.}
	\label{fig:rwithq}
\end{figure}

From Table~\ref{tabel:data_contohlensa} it can be inferred that the ABG has $1.5\times$ smaller values of $\theta_{\infty}$ compared to Bardeen. This means that photon can orbit ABG with smaller radius. While the separation $s$ for the ABG is surprisingly $30\times$ larger, its magnification $r_m$ is smaller than for Bardeen. The NLED thus strengthens the gravitational field by decreasing the innermost distance while at the same time increasing its corresponding separation with the outermost image. Interestingly, the observables in the ABG behave in such opposite ways with the the ones in the Bardeen. In Figs.~\ref{fig:thetainf} it is shown that while the $\theta_\infty$ in Bardeen decreases monotonically, in the ABG case it increases.
From Figs.~\ref{fig:swithq} the separation in the Bardeen model increases monotonically, while in the ABG there exists some maximum value $s=s_{max}$ before which it initially increases and after which it starts decreasing.
Similarly, in Figs.~\ref{fig:rwithq} we see that the $r_m$ decreases unboundedly for the Bardeen as $q$ increases, whereas it decreases to its minimum value before increasing monotonically for the ABG. 

\section{Is Sgr A* a nonsingular black hole?}
\label{sec:shadow}

Having calculated the lensing observables for the Sgr A* if its modeled as the ABG black hole, a tempting question would be: how realistic is Sgr A* as a nonsingular black hole? While to the best of our knowledge we have not found any strong lensing data of Sgr A* to compare with, we do have data for its shadow radius bound. The black hole {\it shadow} is the dark region surrounding it due to the inability of the photon from the background light source to escape the gravitational potential around the black hole. The size of the shadow corresponds to the critical impact parameter of the photon orbit \cite{Stuchlik:2019uvf, Bisnovatyi-Kogan:2017kii,Lu:2019zxb}. The idea that the Sgr A* shadow can be observed was suggested in~\cite{Falcke:1999pj, Melia:2001dy, Kruglov:2020tes}. 
The shadow radius $R_{sh}$ is represented by the critical impact parameter, {\it i.e.,} the impact parameter evaluated at the photon sphere radius $r_{ps}$~\cite{Cunha:2018acu, Perlick:2021aok},
\begin{equation}
	R_{sh} = b_c = r_{ps}\sqrt{\frac{h(r_{ps})}{f(r_{ps})}}.
\end{equation}
For Bardeen with ABG source we will have
\begin{equation}
\label{ABGshadow}
	R_{shABG} = r_{ps}\sqrt{\frac{\left( 1 - \frac{2(6q^2 - r_{ps}^2)}{(q^2 + r_{ps}^2)} \right)^{-1}}{\left(1 - \frac{2mr_{ps}^2}{(r_{ps}^2 + q^2)^{3/2}}\right)}}.
\end{equation}

Recently, Vagnozzi {\it et al}~\cite{Vagnozzi:2022moj} did a comprehensive horizon-scale test using the EHT data from the Sgr A* to constraint a wide range of classical black hole solutions. They calculate $\delta$, the fractional deviation between the inferred and and the Schwarzschild shadow radii, and compared them with the observational estimates by Keck and VLTI (Very Large Telescope Interferometer)~\cite{EventHorizonTelescope:2022xqj}:
\begin{equation}
\delta\simeq-0.060\pm0.065.
\end{equation}
Converting this bound into the shadow radius, and assuming Gaussian uncertainties, the constraint reads
\begin{equation}
4.54\lesssim R_{sh}/M\lesssim5.22,
\end{equation}
and
\begin{equation}
4.20\lesssim R_{sh}/M\lesssim5.56,
\end{equation}
for $1\sigma$ and $2\sigma$ levels, respectively.

\begin{figure}
	\centering
	\includegraphics[scale=1.0]{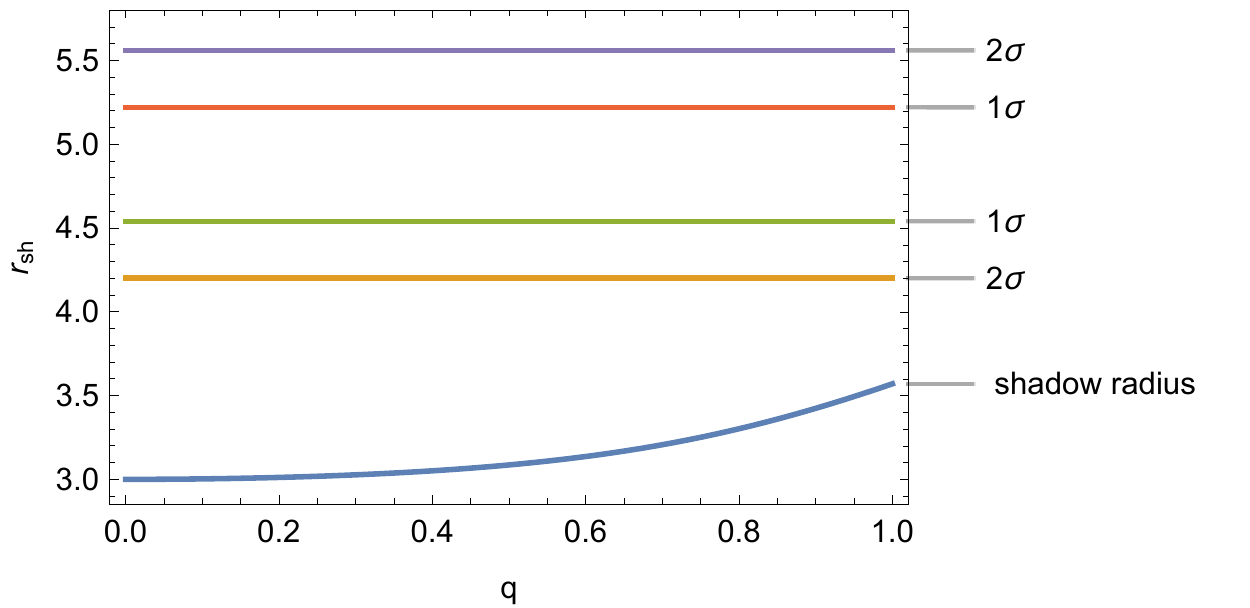}
	\caption{Shadow radius $r_{sh}$ of the ABG black hole as a function of magnetic charge $q$.}
	\label{fig:shadowABG}
\end{figure}

Among the many BH solutions the authors of Ref.~\cite{Vagnozzi:2022moj} scrutinize, they left out the ABG BH unchecked. We plotted the shadow radius as a function of magnetic charge for the ABG BH in Eq.~\eqref{ABGshadow}, , $R_{shABG} =R_{shABG} (q)$. The result is shown in Fig.~\ref{fig:shadowABG}. Interestingly, while the original Bardeen metric passes the horizon-scale test, Fig.~(4) of~\cite{Vagnozzi:2022moj}, the effective metric of ABG does not, as can be seen from the figure. The radius shadow of ABG BH is way below the ETH constraint for Sgr A*. Thus, we conclude that the possibility of Sgr A* being an ABG black hole is ruled out.

\section{Conclusion}
\label{sec:conc}

In this work, we study both the strong lensing and shadow phenomenology of nonsingular ABG black hole. Our approach is different either from~\cite{Eiroa:2010wm, Wei:2015qca} in that we regard the nonsingularity coming from the NLED charge, or from~\cite{Ghaffarnejad:2016dlw} where we used simpler ABG NLED model. 

The NLED reduces photon sphere radius with increasing $q$ until it reaches $q=q_{crit}$, after which it starts to increase monotonically. It also reduces the radius of the photon orbit, and also increases the gravitational field by pulling the innermost distance closer but at the same time stretching its separation distance from the outermost image. Our observable quantities behave differently from the ordinary Bardeen~\cite{Eiroa:2010wm, Wei:2015qca} in terms of the innermost image (Figs.~\ref{fig:thetainf}), the separation between innermost and outermost images (Figs.~\ref{fig:swithq}), and the magnification (Figs.~\ref{fig:rwithq}). Interestingly, the NLED type that we particularly choose to model Bardeen in this paper gives distinct observable results compared to other types. While the model~\cite{Ghaffarnejad:2016dlw} predicts that the angular separation $s$ decreases while the magnification $r_m$ increases with increasing $q$, our results show the opposite. From Table~\ref{tabel:data_contohlensa} we can observe that as the charge increases, the angular separation does increase while the magnification does decrease. 

Lastly, and more importantly, in this work we try to answer a tempting question of whether the Sgr A* at the center of our Milky Way is a nonsingular supermassive BH or not. Based on our shadow radius analysis, we show that Sgr A* cannot be the ABG black hole. Its shadow radius is way below the constraints imposed by EHT observational data. This is, however, is by no means saying that the Sgr A* cannot be nonsingular BH. In Ref.~\cite{Vagnozzi:2022moj} it is shown that the EHT observations are consistent with Sgr A* being a Hayward BH~\cite{Hayward:2005gi}. The static ABG model itself is non-realistic from astrophysical point of view  for Sgr A*, since all black holes are supposed to be rotating. It would be interesting if we can put the rotating nonsingular black holes (for example, see~\cite{Ghosh:2014pba,Amir:2016cen, Abdujabbarov:2016hnw}) to this horizon-scale test to see whether they are suitable to model the Sgr A*. Recently, Walia, Ghosh, and Maharaj tested the three rotating regular BHs (Hayward, Bardeen, and Simpson-Visser) using the EHT observations for Sgr A*~\cite{KumarWalia:2022aop}. They concluded that those three metrics can be still within the EHT bounds and thus the possibility of Sgr A* being one of them cannot be ruled out. However, their black hole metrics are not sourced by the NLED charge. Due to the rotating nature, it is not known whether such exact solutions exist for coupled Einstein-NLED equations.



\acknowledgments

We thank Reyhan Lambaga and Imam Huda for the fruitful discussions on the preliminary stage of this work.

\section*{Data Availability Statement}

Data sharing is not applicable to this article as no data sets were generated or analyzed during the current study.


\end{document}